\documentclass{article}

\usepackage{amsmath,bm}
\usepackage[utf8]{inputenc}
\usepackage[T1]{fontenc}
\usepackage{lmodern} 
\usepackage{graphicx}
\usepackage{gensymb}
\usepackage[euler]{textgreek}
\usepackage[affil-it]{authblk}
\usepackage{tabu}
\usepackage[margin=2.54cm]{geometry}
\usepackage[tiny]{titlesec}
\usepackage{float}
\usepackage{setspace}
\usepackage{caption}
\usepackage[pdftex,
            pdfauthor={S. E. Freeney; J. J. van den Broeke; A. J. J. Harsveld van der Veen; I. Swart; C. Morais Smith},
            pdftitle={Supplementary Material - Edge dependent topology in Kekule lattices},
            pdfsubject={Crystalline topological insulators},
            pdfkeywords={CO on Cu(111), designer quantum matter, STM, topological insulators},
            pdfproducer={},
            pdfcreator={}]{hyperref}

\title{\textbf{Edge-dependent topology in Kekulé lattices}}
\date{\today}
\newcommand*\samethanks[1][\value{footnote}]{\footnotemark[#1]}
\author[1]{S. E. Freeney \thanks{Both authors contributed equally.}}
\author[2]{J. J. van den Broeke  \samethanks}
\author[1]{A. J. J. Harsveld van der Veen}
\author[1]{I. Swart \thanks{Correspondence to: I.Swart@uu.nl, C.deMoraisSmith@uu.nl}}
\author[2]{C. Morais Smith \samethanks}
\affil[1]{Debye Institute for Nanomaterials Science\\ Utrecht University, Netherlands}
\affil[2]{Institute for Theoretical Physics, Utrecht University, Netherlands}

\begin{document}
\vspace*{-1pt}
    {\let\newpage\relax\maketitle}
  \maketitle
  \pagenumbering{arabic}
  \newpage
\captionsetup[figure]{labelfont={bf},labelformat={default},labelsep={colon},name={Fig.}}

\noindent \textbf{Topological states of matter are robust quantum phases, characterised by propagating or
localised edge states in an insulating bulk. Topological boundary states can be triggered by
various mechanisms, for example by strong spin-orbit coupling. In this case, the existence of
topological states does not depend on the termination of the material. On the other hand,
topological phases can also occur in systems without spin-orbit coupling, such as topological
crystalline insulators. In these systems, the protection mechanism originates from the
crystal symmetry. Here, we show that for topological crystalline insulators with the same
bulk, different edge geometries can lead to topological or trivial states. We artificially
engineer and investigate a 2D electronic dimerised honeycomb structure, known as the
Kekulé lattice, on the nanoscale. The surface electrons of Cu(111) are confined into this
geometry by positioning repulsive scatterers (carbon monoxide molecules) with atomic
precision, using the tip of a scanning tunnelling microscope. We show experimentally and
theoretically that for the same bulk, molecular zigzag and partially bearded edges lead to
topological or trivial states in the opposite range of parameters, thus revealing a subtle link
between topology and edge termination. Our results shed further light on the nature of
topological states and might be useful for future manipulations of these states, with the aim
of designing valves or other more complex devices.} \\

\noindent A common assumption concerning topological states of matter is that their existence should
not depend on the sample termination. The quantised conductivity at the edges of the
otherwise insulating material should be insensitive to any detail, except the topology of the
bands. This is indeed the case for the quantum Hall effect [1-3] or for the quantum spin Hall
effect [4-6], which are triggered by a magnetic field or by a strong spin-orbit coupling
respectively, but it does not hold for crystalline topological insulators [7,8]. The reason for
this is that the topological invariant depends on the choice of unit cell, which in turn is
determined by the edge geometry. To establish the relation between edge geometry and
the existence of protected boundary states in topological crystalline insulators
experimentally, it is essential to design lattices that have the required weak and strong
bonds and to have atomically precise edges. \\

\noindent Electrons in engineered potentials can be used as quantum simulators to study the electronic
properties of a large variety of systems, ranging from artificial periodic lattices [9,10] and
quasicrystals [11] to fractals [12]. In these systems, it is possible to control the spin [13] and
orbital degree of freedom [14], as well as the hopping strength between different sites [9,15].
Quantum simulators can be produced by using the tip of a scanning tunnelling microscope
(STM) to manipulate adsorbates to confine electronic states [16] into artificial lattices, or to
manipulate vacancies where electronic states can be contained. The versatility of these types
of artificial lattices is demonstrated by the realisation of topological states of matter.
Vacancies in a chlorine monolayer on Cu(100) have been coupled together to realise
topologically non-trivial domain-wall states in 1D Su Schrieffer Heeger (SSH) chains [17]. In
addition, the manipulation of Fe atoms on the superconducting Re(0001) surface led to the 
realisation of a topological superconductor [18,19]. Recently, the carbon-monoxide (CO) on
Cu(111) platform has been used to create the so-called higher-order topological insulators
(HOTI), in which the topological phase at the boundary exists in at least two dimensions less
than in the bulk. A Kagome lattice has been designed, and the tri-partite nature of the unit cell
was shown to bring further protection to the zero-mode corner states [20]. \\

\noindent Here, we investigate the relation between the emergence of topological states and edge
geometry in a topological crystalline insulator by focusing on the Kekulé lattice. For this lattice,
topologically non-trivial modes should only emerge for certain edge geometries [21]. Fig. 1
shows the Kekulé texture. It can be regarded as a triangular lattice of hexagonal molecules
connected to each other by bonds of strength $t_1$, while the bonds within these hexagonal
molecules have strength $t_0$. We label these nearest-neighbour bonds as inter- ($t_1$) and intra-
($t_0$) hexagon hopping, respectively (see Fig. 1a,b, where $t_0$ is represented by light blue lines and
$t_1$ is represented by navy lines). These bonds, alternated with different strengths, are
reminiscent of the 1D SSH model describing polyacetylene, which is known to exhibit
topological edge modes [22]. Topological edge states occur similarly for the Kekulé lattice:
when a site at the edge is connected only via weak bonds to the rest of the lattice, it forms an
edge mode. This, in addition to sublattice symmetry and mirror symmetry, can give rise to
topologically protected states [21]. \\

\begin{figure}[h!]
    \centering
    \includegraphics[scale=1]{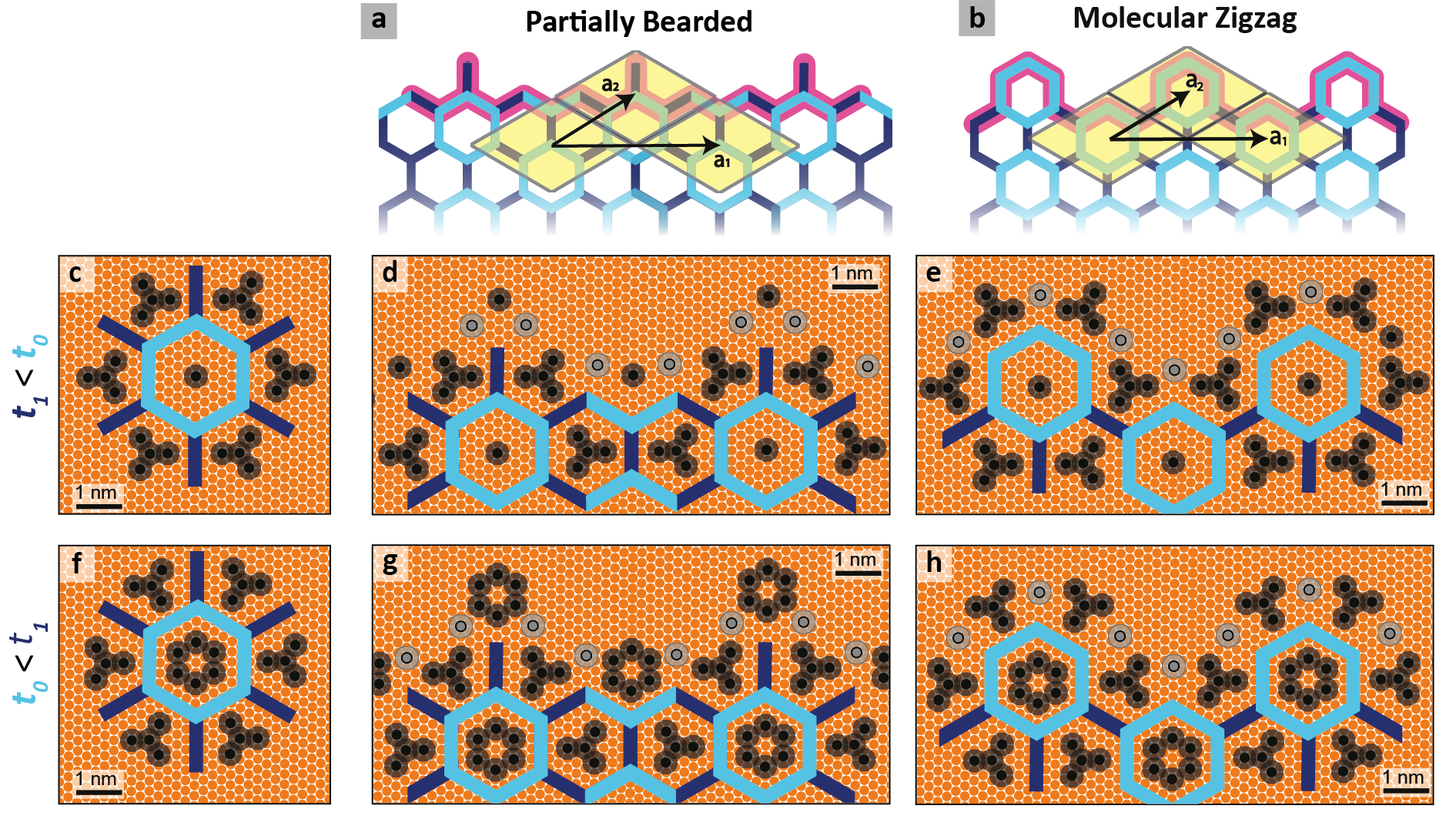}
    \caption{\doublespacing \textbf{Geometry of the Kekulé lattice and the edges investigated.} \textbf{a} and \textbf{b} show the
configuration of bonds at two different edges, partially bearded and molecular zigzag. The
light blue and navy lines indicate the intra- and inter-hexagon hopping parameters
respectively. “Atomic sites” lie at the vertices. The edges are highlighted in pink. The unit cell is
defined by one yellow rhombus. Lattice vectors $\bm{a_1}$ and $\bm{a_2}$ are shown. To form the edge,
translation is performed in a zigzag manner along the $\bm{a_1}$ direction. In row \textbf{c}, the intra-hexagon
hopping is larger than the inter-hexagon hopping ($t_1 < t_0$), while in row \textbf{f} the inverse is true.
The leftmost column shows the positioning of the CO molecules (black dots) for a single
hexagonal unit on the Cu(111) surface, where each orange dot represents a copper atom.
Column \textbf{a} shows the geometry of the partially bearded edge and corresponding configuration
of CO molecules at the lattice terminations on Cu(111) to produce \textbf{d} $t_1 < t_0$ and \textbf{g} $t_0 < t_1$. Column
\textbf{b} shows the same for the molecular zigzag termination, where in \textbf{e} $t_1 < t_0$ and in \textbf{h} $t_0 < t_1$. In \textbf{d},
\textbf{e}, \textbf{g} and \textbf{h}, the grey circles represent additional CO molecules whose purpose is to shield the
edges from the surrounding 2D electron gas.}
\end{figure}

\noindent Theoretically, the Kekulé system has been a benchmark for studies of charge fractionalisation
in the presence of time-reversal symmetry [23]. Moreover, it was proven that these
fractionally charged excitations are semions, hence Abelian anyons carrying $e/2$ charge and
manifesting a $\pi/2$ phase upon braiding [24]. The presence of topological edge states in the
Kekulé lattice was first proposed by Wu et al. [25]. Two phases, one at $t_0 < t_1$ and another at $t_0 > t_1$, separated by a bulk gap closing at the point where the structure is simply a regular
graphene lattice ($t_0 = t_1$), were thought to be analogous to the quantum spin Hall effect as a
consequence of an effective time-reversal symmetry [25]. It was later discovered that this was
not the case, as the presence of edge states does not only depend on the hopping structure,
but also on the edge type, which defines the unit cell. For the two different hopping structures
and the two different types of edge termination, two topological and two trivial phases were
predicted, classified by the mirror winding number [21]. It emerged that the protection of
topological in-gap modes is as a result of the chiral and mirror symmetry of the system [26].
Both these protecting symmetries pose experimental challenges: in realistic finite-size
systems, the mirror symmetry can only be preserved locally, and the usually unavoidable next-nearest neighbour (NNN) hopping breaks the chiral symmetry of the system.
We engineer four finite lattices to experimentally investigate the role of the boundary in
Kekulé systems. To generate these lattices, we use electronic scatterers, in this case CO
molecules, to confine the surface state that manifests as a 2D electron gas on Cu(111). If for
example the scatterers are arranged to form a box, the surface-state electrons confined within
it adopt particle-in-a-box type behaviour, which can be considered analogous to the behaviour
of electrons in an atom. This concept can be taken further; when scatterers are arranged into
an entire lattice, the electrons assume the form of the anti-lattice by sitting between the
scatterers [9]. It is with this principle that we generate the Kekulé lattices with two different
values of hopping parameters. We switch the positioning of the strong and weak bonds (intra- and inter-hexagon), as well as the termination of the structures (bearded or molecular zigzag
[21]). In total, we build four lattices by manipulating up to 522 carbon monoxide atoms per
lattice with atomic precision, using the STM tip. \\

\noindent As a main outcome, we observe that the same Kekulé structure may be trivial or topological,
depending on the termination of the sample. The experimental observations are corroborated
by theoretical calculations using the muffin-tin and tight binding approaches for the specific
experimental realisation, as well as investigations of the underlying crystalline symmetries
protecting the topological phase. \\

\noindent \textbf{Experimental realisation:} The design of the engineered lattices may be observed in Fig. 1a,b.
The leftmost column of the pictographic table in Fig. 1 shows the precise positioning of the CO
molecules on Cu(111) for a single Kekulé unit cell. CO only adsorbs directly on top of surface
atoms of Cu(111). For $t_0 < t_1$, the repulsive potential introduced by the central six CO molecules
serves to diminish the strength of $t_0$ (light blue). In contrast, for $t_1 < t_0$, there is less repulsion
about the single central scatterer [9]. Additionally, for $t_0 < t_1$, each triangularly shaped collection
of four CO molecules reduces the bond strength between hexagons, while for $t_1 < t_0$ they are
rotated 60\degree \: with respect to the opposite design. This allows for a stronger $t_0$ while
simultaneously impinging on the connection between hexagons, decreasing $t_1$. Since the lattice
has triangular symmetry, we have chosen the overall shape of the lattice to be triangular to
allow for the same type of edges on all sides. Symmetry is preserved at the edges, including at
the corners, where there is local resemblance to the edges. The distance between CO
molecules was deliberately chosen. The on-site energy of an electron confined in 2D increases
linearly with the inverse of the area of the boundary containing it. This effect can equivalently
be seen through larger or smaller unit cells effectively leading to n- or p-doping, respectively
[9,14]. Thus, we have chosen a unit cell size that positions the middle of the bulk gap
approximately at the Fermi energy. Grey circles at the edges of the crystal in Fig. 1d,e,g,h
represent CO molecules that were introduced to limit the interaction with the surrounding 2D 
electron gas. Without these additional scatterers, there would be an energy broadening in the
differential conductance spectra/local density of states (LDOS), reducing the energy resolution
so that experimental features would be less clearly defined. The positions of these CO
“blockers” have been carefully chosen, so that the confinement of electrons at the edges is as
equivalent as possible to that of electrons in the bulk, keeping the on-site energy at
approximately $E_F$. This is elaborated upon in the ‘Choice of blocking’ section in the
Supplementary Information. \\

\noindent Two different types of termination have been investigated for each lattice, based on the
theoretical proposal by Kariyado and Hu [21]: the partially bearded edge and the molecular
zigzag edge, as introduced in the first row of the pictographic table in Fig. 1. Below each, the
blueprints for the precise arrangement of the CO molecules used to achieve such edges are
shown for both $t_1 < t_0$ and $t_0 < t_1$. Neither the zigzag nor the armchair edges are included in this
investigation. The zigzag edge has not been created because it leads to well-known results:
due to the breaking of the sublattice symmetry at the zigzag termination, edge states occur
inevitably, and are therefore of no special interest. The lattice terminations presented here
preserve the sublattice symmetry. The armchair terminated configuration has been
theoretically predicted to exhibit gapped edge modes [27]. The HOTI properties of armchair
terminated Kekulé lattices have recently been experimentally investigated in photonic systems
[26]. The protection of the corner modes is however different from the edge modes
investigated here. \\

\begin{figure}[h!]
    \centering
    \includegraphics[scale=1]{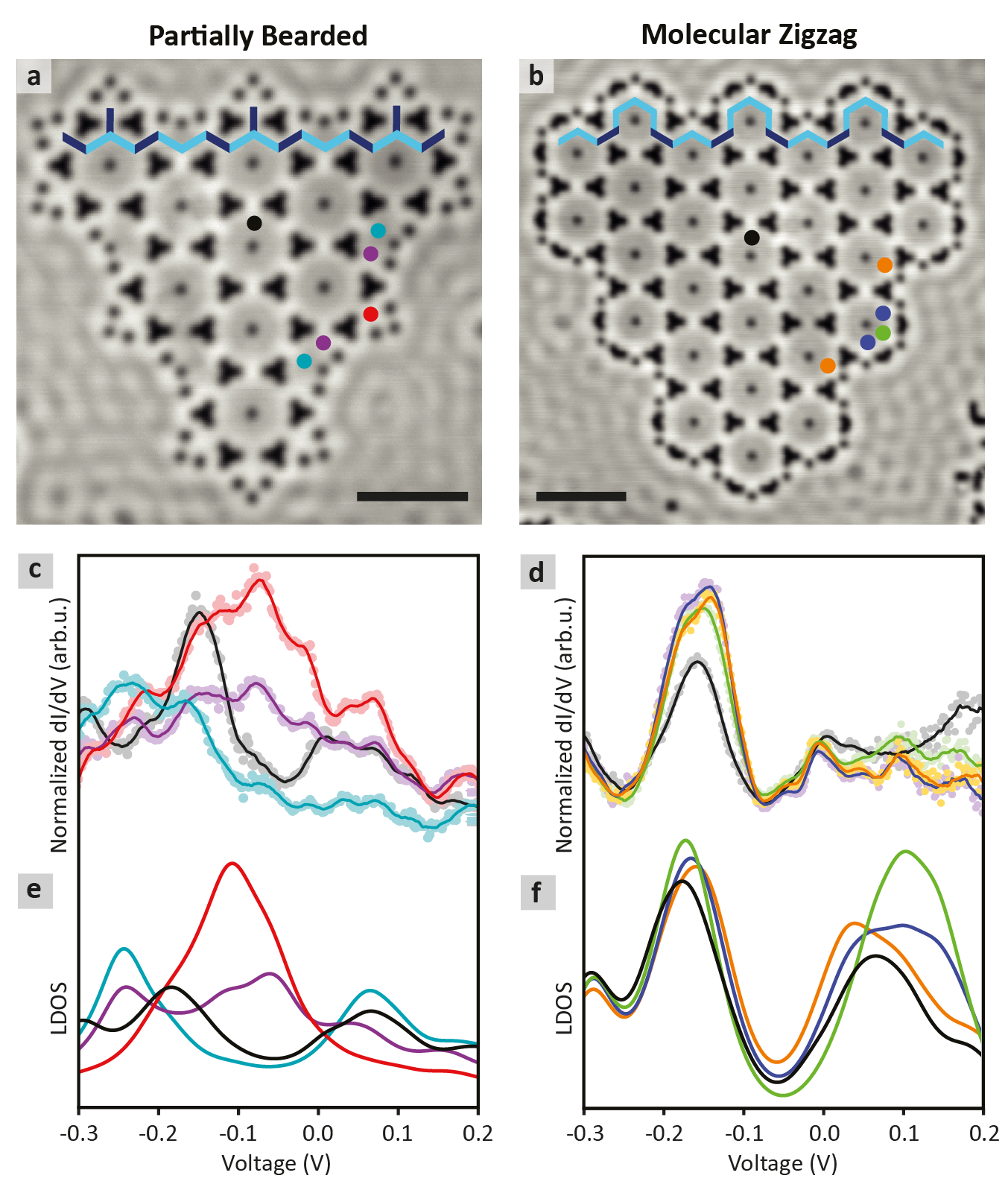}
    \caption{\doublespacing \textbf{Lattices with $\bm{t_1 < t_0}$ and differing edges.} \textbf{a} Topographic image of a partially bearded
edge ($V_{gap}$ = 100 mV, $I_{set}$ = 10 pA); \textbf{b} sample terminated by a molecular zigzag edge ($V_{gap}$ = 100
mV, $I_{set}$ = 100 pA). Scale bars (black) are 5 nm. The corresponding dI/dV spectra normalised by
the ones measured in a region of clean Cu(111) are shown in \textbf{c} and \textbf{d}, and tight binding
calculations are shown in \textbf{e} and \textbf{f}, respectively. The points in the background of the
experimental spectra are the measured values, and the lines overlaid are the moving average.
The colour code indicates the position represented by the dots in \textbf{a} and \textbf{b}. One observes in \textbf{c}
and \textbf{e} that there is an edge mode (red curve) lying at the gap of the bulk (black curve), whereas
in \textbf{d} and \textbf{f} the spectral weight in the bulk and at the edges are identical.}
\end{figure}

\begin{figure}[h!]
    \centering
    \includegraphics[scale=1]{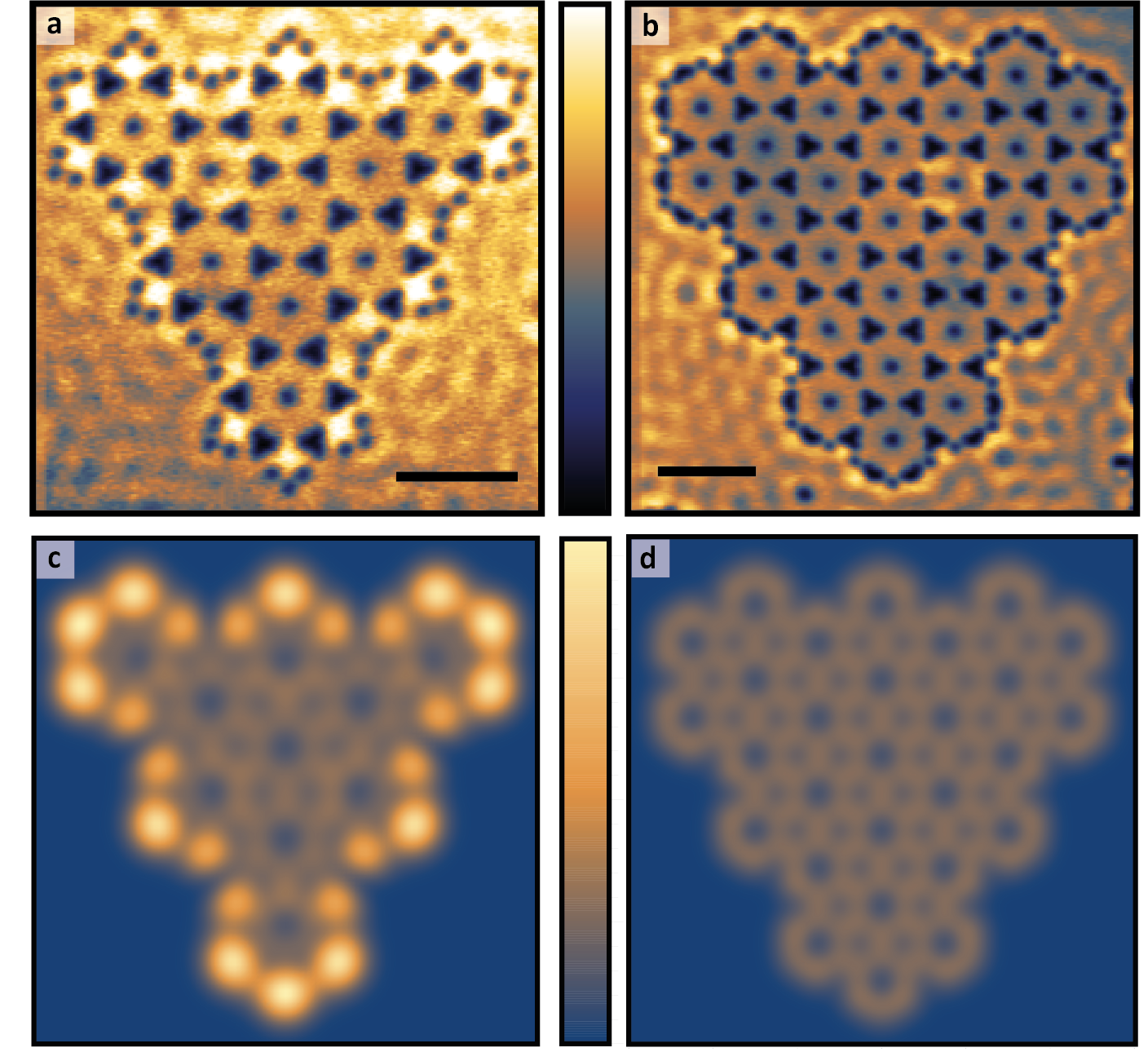}
    \caption{\doublespacing \textbf{LDOS maps for the case $\bm{t_1 < t_0$}.} \textbf{a}, \textbf{b} Differential conductance maps observed
experimentally and \textbf{c}, \textbf{d} the corresponding local density of states calculated theoretically with
tight binding for both types of termination, in the case when the bulk is identical for the two
lattices and $t_1 < t_0$. Experimental maps were acquired at -65 mV and local density of states
calculated at -65 meV. Scale bars (black) are 5 nm in length. Brighter pixels represent higher
LDOS. The lattice terminated by a partially bearded edge (\textbf{a}, \textbf{c}) exhibits edge states, whereas
the molecular zigzag terminated lattice does not (\textbf{b}, \textbf{d}).}
\end{figure}

\noindent \textbf{Results:} Two lattices with $t_1 < t_0$ are shown in the constant-current topographic images in Fig.
2a,b. They have the same bulk but differ in the geometry of their edges. Differential
conductance spectroscopy was performed at certain positions in the bulk and at the edge,
represented by the coloured dots overlaying the topographic images. The corresponding
spectra are given in Fig. 2c,d. We observe that whereas the dI/dV spectra taken on the
molecule-zigzag terminated lattice show the same features at the bulk and edge, there is a
stark difference in the appearance of the LDOS specta at the bulk and edge for the partially
bearded edged lattice. The spectrum of bulk sites (indicated in black) shows two peaks
associated with the valence (at -0.15 V) and conduction bands (+ 0.05 V), separated by a gap.
In contrast, the spectrum of the site indicated in red shows a large peak positioned at the
energy of the bulk gap. There is a moderately elevated density of states localised at purple
sites compared to the bulk, and the cyan sites have a DOS similar to the bulk. The
experimentally observed features are also found in the local density of states, as calculated by
tight binding (see Fig. 2e,f) and muffin-tin simulations (see section ‘Muffin-tin calculations’ in
the Supplementary Information). These results, both theoretical and experimental, support
the theoretical prediction based on calculation of the topological invariant (the mirror winding
number) [21] that the edge mode at the partially bearded edge is topological when $t_1 < t_0$. We
also show in the Supplementary Information section entitled ‘Defects at the edges’ that the
introduction of CO molecules to act as edge defects has no influence on the existence of the
edge states. The resilience of edge states to defects is a signature of topological states.
This conclusion is further supported by the differential conductance maps shown in Fig. 3. The
maps were acquired at an energy corresponding to the approximate centre of the bulk gap. In
Fig. 3a, one observes a well-defined edge mode at the bearded edge (the brighter the pixel,
the higher the LDOS), whereas the bulk is insulating. On the other hand, in Fig. 3b there is no 
9
difference between the spectral weight in the bulk and at the molecular zigzag edge. The
corresponding LDOS evaluated theoretically from a tight binding model display a very good
agreement with the experiments (see Fig. 3c,d.) 

\begin{figure}[h!]
    \centering
    \includegraphics[scale=1]{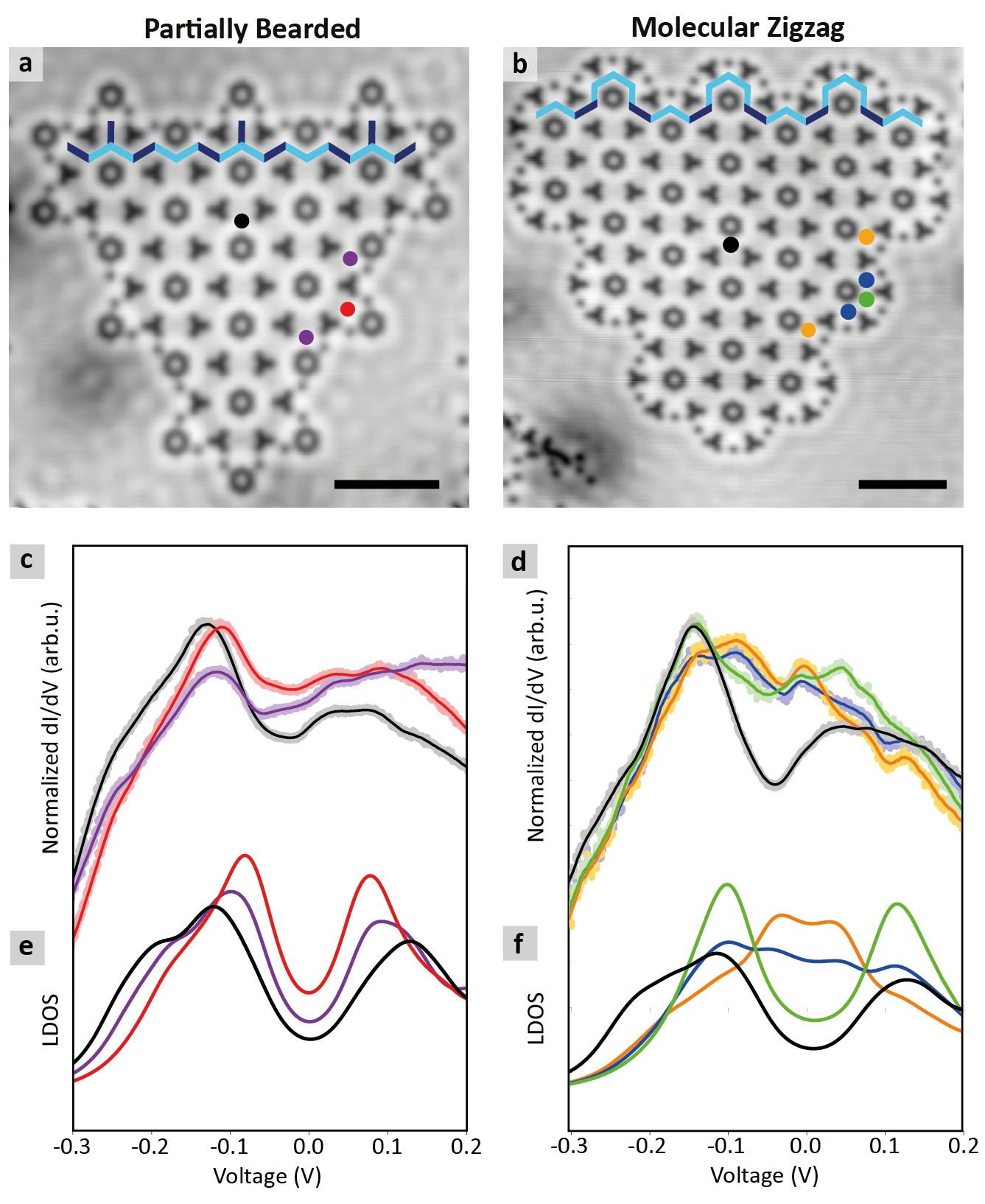}
    \caption{\doublespacing \textbf{Lattices with $\bm{t_0 < t_1}$ and differing edges.} Topographs are shown for \textbf{a} the partially
bearded edge and \textbf{b} the molecular zigzag edge. Both scans were acquired using $V_{gap}$ = 100 mV
and $I_{set}$ = 30 pA. Scale bars (black) are 5 nm. Experimental spectra were averaged with spectra
of the clean Cu(111). The points in the background represent the actual measured values and
the lines overlaying them are the moving averages. \textbf{c} For the partially bearded termination, the
spectrum is identical at the bulk and edges. \textbf{d} For the molecular zigzag edge there is a
heightened LDOS at several positions at the edge. The tight binding calculated spectra \textbf{e}, \textbf{f}
behave similarly. The behaviour is precisely opposite to that shown in Fig. 2. }
\end{figure}

\noindent The situation is reversed when the hopping strengths are inverted [21]; Fig. 4 shows
topographs and LDOS spectra for the Kekulé lattice in the opposite regime, where $t_0 < t_1$. In
this case, for the partially bearded edged lattice (topograph shown in Fig. 4a), the
experimental spectra at different edge positions follow the behaviour of the spectrum in the
bulk of the crystal (Fig. 4c): there is a dip in the experimentally acquired LDOS around -20mV
for all positions measured which implies trivially insulating behaviour throughout. The tight
binding calculated LDOS in Fig. 4e is in agreement with experiment. At the molecular zigzag
edge (topograph in Fig. 4b) for the same $t_0 < t_1$ case, there is a markedly higher LDOS at the
edge positions at energies corresponding to the bulk gap, (see experimental data in Fig. 4d
tight binding in Fig. 4f). The spectrum taken at the position marked in green in Fig. 4d does
not show a minimum as pronounced as in the tight binding spectrum in Fig. 4f. Additionally,
the blue and orange spectra are quite similar to each other. A slight difference between
experiment and theory is also observed in Fig. 4c,e, where the peak corresponding to the
conduction band is less defined for all spectra in experiment compared to theory. The
difference can likely be attributed to the specific configuration of the STM tip apex used for
taking spectra on these lattices. As a general statement, the DOS of the tip always makes a
contribution to a LDOS measurement, and in this case it may have affected the spectra
(even after the normalisation with the Cu(111) background). Another possibility is that the
tip was asymmetric, leading to a slight mismatch between the desired position for
spectroscopy and the actual position. A varying position at which LDOS spectra are acquired
can influence the shape of the spectra. \\

\noindent From these results, we can assert that for $t_0 < t_1$ the sample terminated by a molecule zigzag
edge exhibits pronounced edge states, while the sample terminated by a partially bearded
edge has no edge states. The conclusions are further illustrated by experimental differential
conductance maps (Fig. 5a,b) and LDOS maps calculated with tight binding (Fig 5.c,d). For the
partially bearded terminated lattice, the LDOS (given by pixel intensity) is approximately
equivalent at the edges and the bulk. Intensity is distinctly higher at the molecular zigzag edge
compared to the bulk of the lattice. This is the reverse outcome as for the opposite hopping
parameter regime. \\

\begin{figure}[h!]
    \centering
    \includegraphics[scale=1]{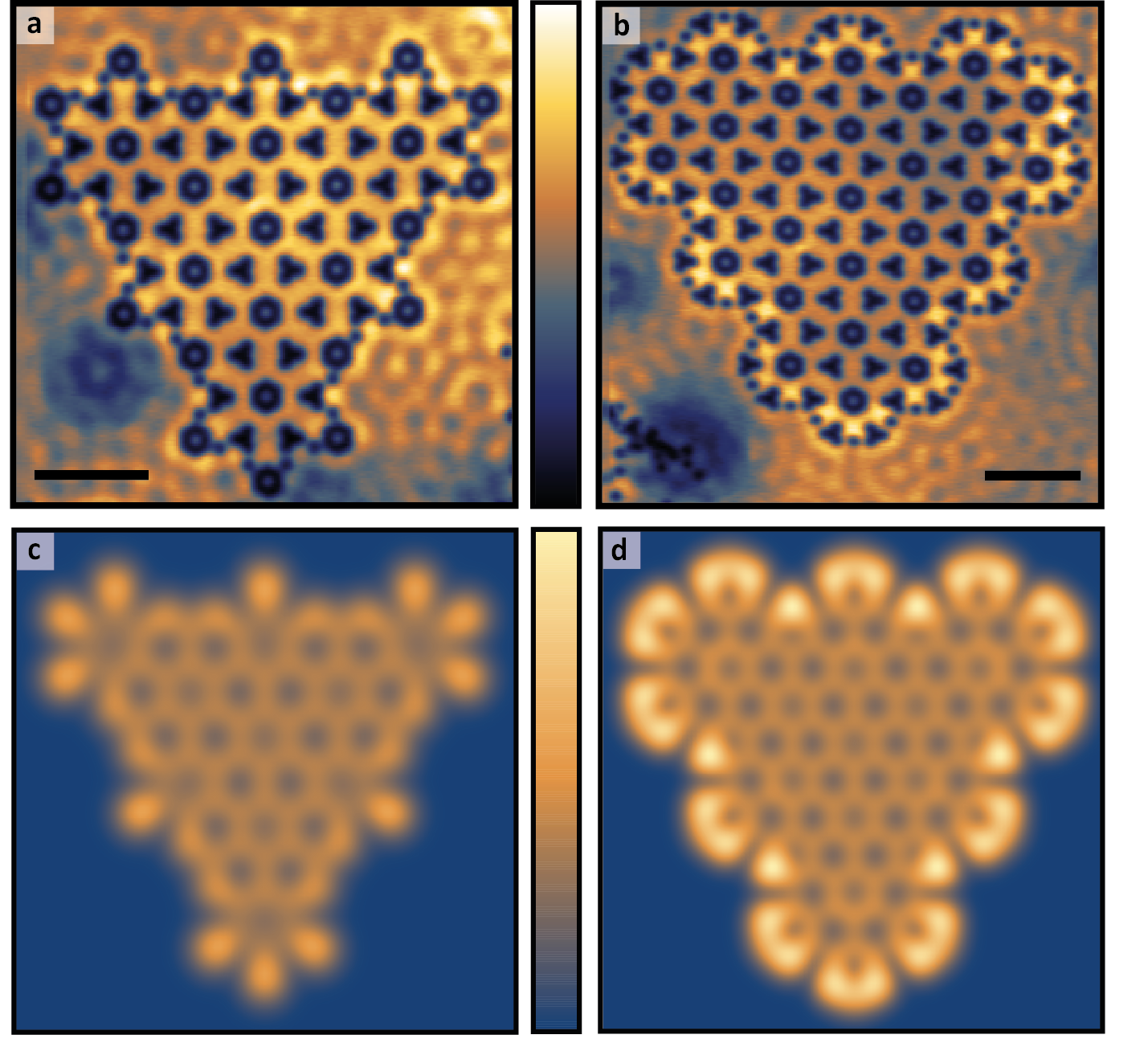}
    \caption{\doublespacing \textbf{LDOS maps for the case $\bm{t_0 < t_1$}.} Differential conductance maps and LDOS for the
configurations shown in Fig. 4. Images \textbf{a}, \textbf{b} are experimentally acquired and \textbf{c}, \textbf{d} have been
calculated using tight binding. The colour scale represents the LDOS; the brighter the colour,
the higher the LDOS. Now, the edge states can be observed for the molecular zigzag edge, and
no longer for the partially bearded one. Experimental maps were acquired at -20 mV and local
density of states calculated at -20 meV. Scale bars (black) represent 5 nm.}
\end{figure}

\noindent To find the appropriate hopping parameters for use in tight binding ($t_1 = 0.7 \: t_0$ for the models in
Fig. 2, 3 and $t_0 = 0.67 \: t_1$ for the models in Fig. 4, 5), the tight binding calculated band structures
were matched to those calculated using the muffin-tin method. Besides the hopping
parameters $t_0$ and $t_1$, orbital overlap and NNN hopping between and within hexagons were
considered in making the fit. The full list of obtained parameter values and a detailed
description of the matching procedure and muffin-tin calculations is given in the
Supplementary Information section ‘Muffin-tin calculations’. Although the orbital overlap
deforms the band structure and is therefore of vital importance to fully understand the
experimental results, it was numerically verified that it does not break the topological
protection of the edge states in the Kekulé lattice. The NNN hopping however, breaks chiral
symmetry. It was found that most NNN hopping parameters were rather small ($\leq 0.02 \:  t_0$) due
to the clustered CO structure. Only the NNN hopping within the hexagon for the $t_1 < t_0$ design
($0.2 \: t_0$) is larger, as there is only one CO in the middle of the hexagons. Therefore we expect
that the chiral symmetry is weakly perturbed for this case. \\

\noindent \textbf{Finite-size effects:} To understand the impact of finite-size effects, we first study how edge
states are protected in the ribbon geometry, and then investigate how these features change
for the finite structures built experimentally. Kariyado et al. [21] found that the mirror winding
number protects the zero energy crossing of the edge modes in the Kekulé system. As the
calculation of this invariant requires both chiral symmetry and reflection symmetry $M_y$ along
the line passing through two directly opposite sites in a hexagon, both symmetries need to be
present to protect the edge states. This has been confirmed by Noh et al. [26] by numerically
adding perturbations to the Hamiltonian in the case of armchair terminated Kekulé lattices.
The $M_y$ symmetry is broken and the edge modes become gapped.

\noindent When a system can be divided in two subsystems that only couple to each other and never to
themselves, the system possesses chiral symmetry. The chiral symmetry leads to a spectrum
that is symmetric around zero energy. This means that zero modes can only move away from
zero energy in pairs. If there are more sites of one subsystem than of the other on the edge,
but not in the rest of the structure, this can result in zero modes on the edge, as in graphene
ribbons with a zigzag termination [28,29]. The edge geometry considered here contains
equally many sites of each sublattice. Thus, chiral symmetry alone does not enforce the
existence of edge states. To understand the protection of zero modes in the system, we should
therefore also consider the reflection symmetry $M_y$. At the gamma point in the Brillouin zone,
$M_y$ commutes with the Hamiltonian. Hence, the Hamiltonian needs to have the same
eigenstates as $M_y$, and states which are even and odd under $M_y$ cannot mix. This mechanism
can prevent two zero modes on the edge of a Kekulé ribbon to mix, thus pinning them at zero
energy due to the chiral symmetry. 

\noindent The Kekulé lattices realised here have (approximate) chiral symmetry, since the NNN hopping
is relatively small. The $M_y$ symmetry is preserved locally. In the experimental designs, the 
12
lattice sites are locally affected by the same environment as they would be in an infinitely long
ribbon, as illustrated in Fig 6a. However, the global mirror symmetry present in the ribbon is
broken in the finite lattice: the boundary is not fully periodic due to modulations to form the
corner. Moreover, the lattice is relatively small; thus the momenta are not continuous and a
state with zero momentum (the Γ point) does not need to exist. We show the evolution of the
energy levels upon tuning the ratio $t_0/t_1$ for the molecular zigzag terminated lattice in Fig. 6
b,c. One observes that upon increasing system size, the resolution in the discretisation
procedure increases. While the zero modes already deviate from 0 energy at roughly $0.7 \: t_0/t_1$
in Fig. 6b, in Fig. 6c, they remain close to 0 up to a larger value of $t_0/t_1$, about $0.9 \: t_0/t_1$. \\

\begin{figure}[h!]
    \centering
    \includegraphics[scale=0.9]{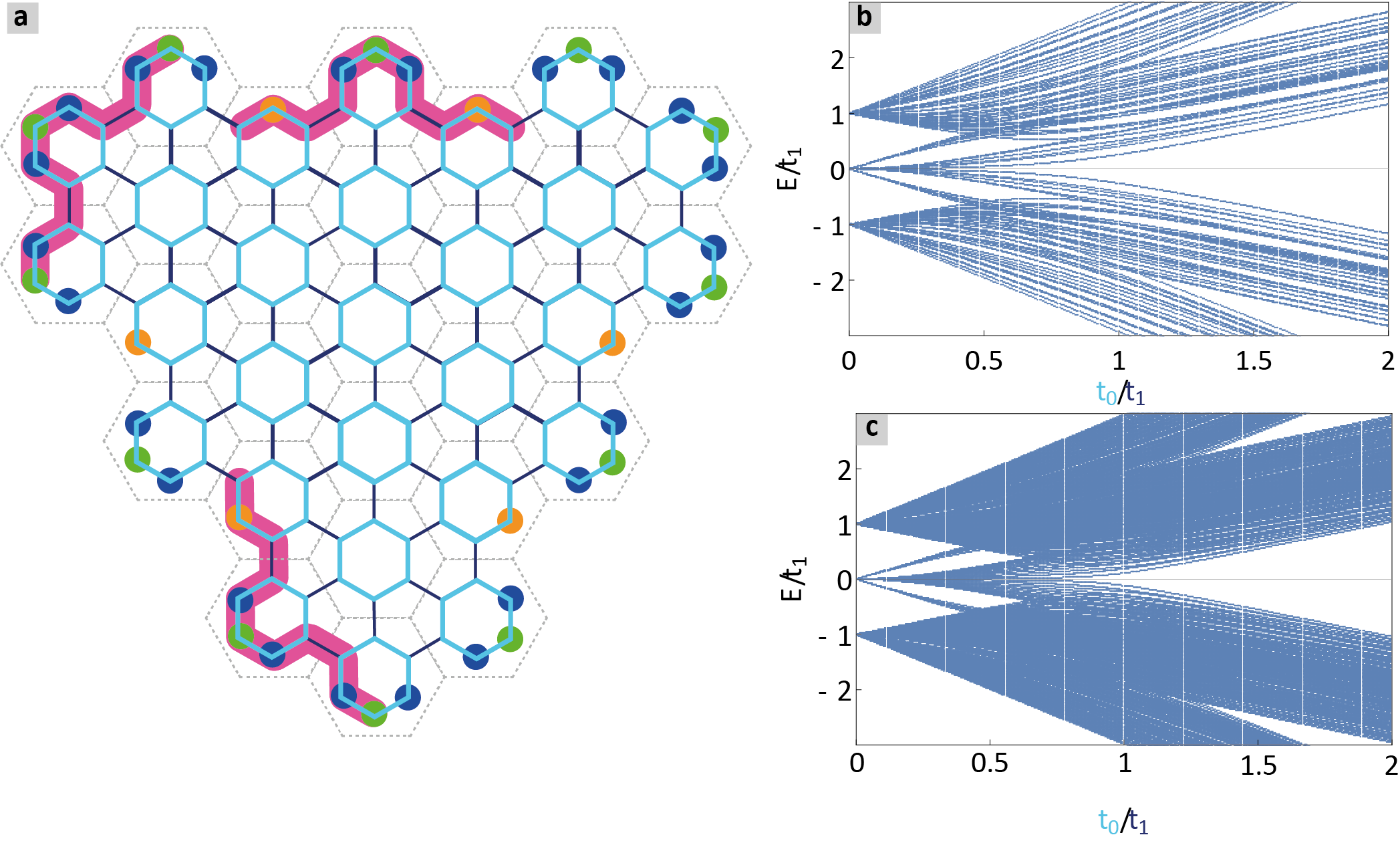}
    \caption{\doublespacing \textbf{Finite size effects.} \textbf{a} Illustration of the finite molecular-zigzag terminated lattice.
Green represents protruding sites that couple weakly to two blue sites, orange represents
sites sitting in a “cove” at the edge of the lattice. The sections shown in pink have the same
local environment. \textbf{b}, \textbf{c} Energy spectrum as a function of $t_0/t_1$. The spectrum is shown for the
system size used in the experiments with 28 hexagons in total (\textbf{b}) and for a theoretical
structure that has the same corner and edge, but contains 163 hexagons (\textbf{c}).}
\end{figure}

\newpage
\noindent \textbf{Discussion and Outlook:} The detection of edge modes in this finite-size system is surprising. In
translationally invariant ribbons, the mirror and chiral symmetries pin the edge modes to zero
energy at the Γ point in the Brillouin zone. However, here we investigate a finite and relatively
small system, without translational symmetry and for which a Brillouin zone cannot be
defined. Furthermore, the chiral symmetry is broken in the $t_1 < t_0$ regime due to a significant
NNN hopping, and the mirror symmetry is not globally preserved, as in the ribbon geometry
that was used for the theoretical predictions [21]. This indicates that the edge modes are
remarkably robust to weak symmetry breaking and finite-size effects.  \\

Our experimental observations, which are corroborated by theoretical calculations in the
continuum (muffin-tin) and in a discrete lattice (tight binding model) confirm that the
existence of a topological phase in symmetry protected topological insulators is a subtle issue.
It does not depend uniquely on the form of the bulk, and sets a boundary of validity to a naïve
interpretation of the bulk-boundary correspondence. Our results indicate that devices made
by the same bulk, in which the termination is adjusted accordingly, could be used to create
valves and manipulate the edge modes at will. Since all the results presented here are generic,
they could be promptly transferred to other kinds of condensed-matter setups, in
semiconductors or metallic surfaces, thus extending our findings to other kinds of electronic
systems.

\newpage
\section*{References}
[1] Von Klitzing, K. The quantized Hall effect. \textit{Rev. Mod. Phys.} \textbf{58}, 519 (1986). \\

\noindent [2] Thouless, D. J., Kohmoto, M., Nightingale, M. P., \& den Nijs, M. Quantized Hall conductance
in a two-dimensional periodic potential. \textit{Phys. Rev. Lett.} \textbf{49}, 405 (1982).\\

\noindent [3] Altland, A., \& Zirnbauer, M. R. Nonstandard symmetry classes in mesoscopic normalsuperconducting hybrid structures. \textit{Phys. Rev. B} \textbf{55}, 1142 (1997).\\

\noindent [4] Kane, C. L., \& Mele, E. J. Quantum spin Hall effect in graphene. \textit{Phys. Rev. Lett.} \textbf{95}, 226801
(2005).\\

\noindent [5] Bernevig, B. A., Hughes, T. L., \& Zhang, S. C. Quantum spin Hall effect and topological phase
transition in HgTe quantum wells. \textit{Science} \textbf{314}, 1757-1761 (2006).\\

\noindent [6] König, M., Wiedmann, S., Brüne, C., Roth, A., Buhmann, H., Molenkamp, L.W., Qi, X.L. and
Zhang, S.C., Quantum spin Hall insulator state in HgTe quantum wells. \textit{Science} \textbf{318}, 766-770
(2007).\\

\noindent [7] Fu, L., Topological crystalline insulators. \textit{Phys. Rev. Lett.} \textbf{106}, 106802 (2011).\\

\noindent [8] Slager, R. J., Mesaros, A., Juričić, V., \& Zaanen, J. The space group classification of
topological band-insulators. \textit{Nat. Phys.} \textbf{9}, 98 (2013). \\

\noindent [9] Gomes, K. K., Mar, W., Ko, W., Guinea, F. \& Manoharan, H. C. Designer Dirac fermions and
topological phases in molecular graphene. \textit{Nature} \textbf{483}, 306–310 (2012).\\

\noindent [10] Slot, M.R., Gardenier, T.S., Jacobse, P.H., van Miert, G.C., Kempkes, S.N., Zevenhuizen, S.J.,
Smith, C.M., Vanmaekelbergh, D. and Swart, I. Experimental realization and characterization of
an electronic Lieb lattice. \textit{Nat. Phys.} \textbf{13}, 672 (2017).\\

\noindent [11] Collins, L. C., Witte, T. G., Silverman, R., Green, D. B. \& Gomes, K. K. Imaging quasiperiodic
electronic states in a synthetic Penrose tiling. \textit{Nat. Commun.} \textbf{8}, 15961 (2017).\\

\noindent [12] Kempkes, S.N., Slot, M.R., Freeney, S.E., Zevenhuizen, S.J.M., Vanmaekelbergh, D., Swart,
I. and Smith, C.M. Design and characterization of electrons in a fractal geometry. \textit{Nat. Phys.} \textbf{15},
127 (2019).\\

\noindent [13] Figgins, J., Mattos, L. S., Mar, W., Chen, Y. T., Manoharan, H. C., \& Morr, D. K. Quantum
Engineered Kondo Lattices. \textit{arXiv preprint} arXiv:1902.04680 (2019).\\

\noindent [14] Slot, M.R., Kempkes, S.N., Knol, E.J., van Weerdenburg, W.M.J., van den Broeke, J.J.,
Wegner, D., Vanmaekelbergh, D., Khajetoorians, A.A., Smith, C.M. and Swart, I. p-band
engineering in artificial electronic lattices. \textit{Phys. Rev. X.} \textbf{9}, 011009 (2019).\\

\noindent [15] Girovsky, J., Lado, J., Kalff, F., Fahrenfort, E., Peters, L., Fernández-Rossier, J. and Otte, S.,
2017. Emergence of quasiparticle Bloch states in artificial crystals crafted atom-by-atom.
\textit{SciPost Phys.} \textbf{2}, 20 (2017).
\\

\noindent [16] Crommie, M. F., Lutz, C. P. \& Eigler, D. M. Confinement of electrons to quantum corrals on
a metal surface. \textit{Science} \textbf{262}, 218–220 (1993). \\

\noindent [17] Drost, R., Ojanen, T., Harju, A. \& Liljeroth, P. Topological states in engineered atomic
lattices. \textit{Nat. Phys.} \textbf{13}, 668–671 (2017).\\

\noindent [18] Kim, H., Palacio-Morales, A., Posske, T., Rózsa, L., Palotás, K., Szunyogh, L., Thorwart, M.
and Wiesendanger, R. Toward tailoring Majorana bound states in artificially constructed
magnetic atom chains on elemental superconductors. \textit{Sci. Adv.} \textbf{4}, 5251 (2018).\\

\noindent [19] Kamlapure, A., Cornils, L., Wiebe, J., \& Wiesendanger, R. Engineering the spin couplings in
atomically crafted spin chains on an elemental superconductor. \textit{Nat. Commun.} \textbf{9}, 3253 (2018).\\

\noindent [20] Kempkes, S. N. et al. Robust zero-energy modes in an electronic higher-order topological
insulator: the dimerized Kagome lattice. \textit{arXiv preprint} arXiv:1905.06053(2019).\\

\noindent [21] Kariyado, T., \& Hu, X. Topological states characterized by mirror winding numbers in
graphene with bond modulation. \textit{Sci. Rep.} \textbf{7}, 16515 (2017).\\

\noindent [22] Heeger, A. J., Kivelson S., Schrieffer, J. R., Su, W. -P. Solitons in conducting polymers. \textit{Rev.
Mod. Phys.} \textbf{60}, 781–850 (1988).\\

\noindent [23] Hou, Chang-Yu, Chamon, C., \& Mudry, C. Electron Fractionalization in Two-Dimensional
Graphenelike Structures. \textit{Phys. Rev. Lett.} \textbf{98}, 186809 (2007).\\

\noindent [24] Seradjeh, B. \& Franz, M. Fractional statistics of topological defects in graphene and
related structures. \textit{Phys. Rev. Lett.} \textbf{101}, 146401 (2008).\\

\noindent [25] Wu, L. H., \& Hu, X. Topological properties of electrons in honeycomb lattice with detuned
hopping energy. \textit{Sci. Rep.} \textbf{6}, 24347 (2016).\\

\noindent [26] Noh, J., Benalcazar, W. A., Huang, S., Collins, M. J., Chen, K. P., Hughes, T. L., \& Rechtsman,
M. C. Topological protection of photonic mid-gap defect modes. \textit{Nat. Photonics} \textbf{12}, 408 (2018).\\

\noindent [27] Liu, F., Yamamoto, M., \& Wakabayashi, K. Topological edge states of honeycomb lattices
with zero berry curvature. \textit{J. Phys. Soc. Jpn.} \textbf{86}, 123707 (2017).\\

\noindent [28] Nakada, K., Fujita, M., Dresselhaus, G. and Dresselhaus, M.S. Edge state in graphene
ribbons: Nanometer size effect and edge shape dependence. \textit{Phys. Rev. B.} \textbf{54} 17954 (1996)\\

\noindent [29] Fujita, M., Wakabayashi, K., Nakada, K., Kusakabe, K. Peculiar localized state at zigzag
graphite edge. \textit{J. Phys. Soc. Jpn.} \textbf{65}, 1920 (1996)\\

\noindent [30] Nečas, D. and Klapetek, P. Gwyddion: an open-source software for SPM data analysis.
\textit{Open Physics} \textbf{10}, 181 (2012).\\

\noindent [31] Li, S., Qiu, W. X., \& Gao, J. H. Designing artificial two dimensional electron lattice on metal
surface: a Kagome-like lattice as an example. \textit{Nanoscale}, \textbf{8}, 12747-12754 (2016). \\

\newpage
\section*{METHODS}

\noindent \textbf{Experiment:} Atomic manipulation, scanning tunnelling microscopy and spectroscopy were
performed using a commercially available Scienta Omicron LT-STM. A Cu(111) surface was
prepared to atomic flatness by repeated cycles of sputtering with Ar$^+$ and annealing at
approximately 550\degree C. Carbon monoxide was then deposited onto the Cu(111) surface within
the cooled microscope head at a pressure of $1.3\times 10^{-8}$ mbar for 1 minute to achieve a coverage
of roughly 0.5 CO molecules per nm$^2$. Following this, the microscope head was kept at
constant UHV (in the range of $10^{-11}$ mbar) and at a temperature of 4.5 K during construction of
the lattices and measurements. An STM tip was cut from platinum-iridium wire, which was
conditioned in-situ by repeatedly dipping the tip into the surface and/or applying voltage
pulses between tip and sample. This leaves the tip with a randomly shaped apex made from
copper atoms, and the process was considered complete when the tip satisfactorily performed
the desired task (either atom manipulation, imaging or spectroscopy). STM topographs were
acquired in constant current mode. Plane subtraction was performed on the topographs. Atom 
manipulation was performed with a bias voltage of 20 mV and constant current maintained
with a feedback loop ranging from 10 nA to 60 nA depending on the condition of the tip.
Differential conductance spectra and maps were acquired in constant height mode with bias
modulation provided by a lock-in amplifier. The amplitude of the modulation was 10 mV r.m.s
at a frequency of 273 Hz. Integration time for signal acquisition was 50 ms at the lock-in
amplifier for spectra and 20 ms for each pixel in the differential conductance maps. The
differential conductance spectra were processed by averaging over numerous sites of
equivalent type or repeated measurements within the same site, then dividing this average by
the average of many spectra on bare Cu(111). The purpose of the division by spectra on bare
Cu(111) is to eliminate LDOS contributions from the tip and from the copper itself. Processing
of the differential conductance maps included alignment of the forward and backwards scans,
then averaging the two. A small amount of Gaussian blurring was applied to reduce the
appearance of noise in each map, except for the trivial partially bearded edged lattice, for
which this was not necessary. The “sky” colour map, which is perceptually uniform, was used
from the freely available open source program Gwyddion [30] (with which all processing was
performed). \\

\noindent \textbf{Theory:} Besides the two types of nearest-neighbour hopping (inter- and intra- hexagon
hopping), the finite size tight binding calculations included orbital overlap and next-nearest
neighbour (NNN) hopping. To obtain the tight binding parameters, the periodic model was fit
to a muffin-tin band structure. For muffin-tin calculations, the experimental setup of CO on
Cu(111) is modelled by describing the CO molecules as disk shaped protrusions in an otherwise
constant two dimensional potential landscape. Here, a disk diameter of 0.6 nm and a potential
height of 0.9 eV was used as in ref. [11]. To obtain the band structure, the corresponding 
Hamiltonian was expanded (up to the 5th Fourier component) in the plane-wave basis. This
was then solved numerically using the analytically known Fourier components of the muffintin potential [31]. In order to investigate the optimal blocking, the time independent
Schrödinger equation was solved using the NDEigensystem routine of Mathematica for the
muffin-tin potential of several different designs. Here, von Neumann boundary conditions
were used. In order to compare the finite size results of both the muffin-tin and tight binding
calculations to experiments, a broadening of 0.08 eV originating from the scattering of
electrons by the bulk due to the presence of CO molecules was included. LDOS maps were
obtained from the tight binding eigenvectors $\Psi_\epsilon$ with energy $\epsilon$ according to:

\begin{equation}
  LDOS(x,y,\epsilon) = \sum_{\epsilon'} \vert \sum_{i} exp(-x-x_i)^2 1.15a) \Psi_{i,\epsilon}(x,y) \vert ^2 L(\epsilon - \epsilon') ,
\end{equation}

where $a$ is the lattice constant of the Kekulé lattice, $L$ is the Lorentzian broadening function,
$i$ enumerates the sites and $x_i$ is the position of site $i$. Further detail regarding the model used
here is given in the ‘Tight binding description’ section of the Supplementary Information.\\

\newpage
\noindent \textbf{Supplementary Information} is available in the online version of the article.\\

\noindent \textbf{Acknowledgements:} We would like to acknowledge Marcel Franz for fruitful discussions. IS
and CMS acknowledge funding from NWO via grants 16PR3245 and DDC13.\\

\noindent \textbf{Author contributions:} SEF performed the experiment and data analysis with contributions
from AJJHvdV under the supervision of IS. JJvdB performed the theoretical calculations
under the supervision of CMS. CMS and SEF wrote the manuscript with input from all
authors. JJvdB and SEF wrote the SI.\\
\vspace{20px}

\end{document}


\vspace*{-1pt}
    {\let\newpage\relax\maketitle}
  \maketitle
  \pagenumbering{arabic}
  \newpage
\captionsetup[figure]{labelfont={bf},labelformat={default},labelsep={colon},name={Supplementary Figure}}

\section*{Tight binding description}
The basic periodic tight binding model of the Kekulé system has only two parameters, the intra- and the inter-hexagonal nearest-neighbour (NN) hopping parameters. However, in order to accurately describe the experimental results, these are not sufficient. 

\noindent The first effect that must be taken into account is the broadening of the peaks in the experimental differential conductance spectra due to scattering or temperature effects. Without broadening, the local density of states (LDOS) $L$ for an energy $\epsilon$ is determined by the wave function $\Psi$ and given by

\begin{equation} \label{eq:1}
  L(x,y,\epsilon) = \sum_{i} \vert \Psi_{\epsilon_{i}}(x,y) \vert ^2 \delta(\epsilon - \epsilon_i) . \tag{Supplementary Equation 1}
\end{equation}

\noindent For the artificial lattices that we investigate here, the main contribution to broadening comes from the presence of the CO molecules on the copper surface, causing electrons to scatter from the surface state into the bulk. This can be described by replacing the delta function in Supplementary Equation 1 by a Lorentzian. The LDOS in now described by  
\begin{equation} \label{eq:2}
  L(x,y,\epsilon) = \sum_{i} \vert \Psi_{\epsilon_{i}}(x,y) \vert ^2 \frac{b}{(\epsilon - \epsilon_i)^2 + (\frac{b}{2})^2} . \tag{Supplementary Equation 2}
\end{equation}

\noindent where $b$ = 0.08 eV in this setup. 
A second significant effect is orbital overlap, which describes the non-zero overlap between the orbitals of neighbouring sites. This leads to the generalised eigenvalue equation $H\Psi = ES \Psi$, where $H$ is the Hamiltonian, $E$ is the energy, and $S$ is the overlap matrix. In order to limit the number of parameters to match and to avoid overfitting, we only considered NN overlap. Thus, there are two orbital overlap parameters for each design: the orbital overlap of two sites in the same hexagon, and overlap between sites of two different hexagons. If all orbitals are orthogonal, the generalised eigenvalue equation reduces to the standard eigenvalue equation, $H\Psi = E \Psi$.

\noindent Finally, next-nearest-neighbour (NNN) hopping must also be included. This again gives two extra parameters, NNN hopping within and between hexagons. The magnitude of both parameters were greatly reduced by the use of clustered CO molecules. In the $t_0 > t_1$ design, however, a single CO molecule is used in the centre of the hexagon, resulting in an intra-hexagon NNN hopping of 0.2 $t_0$. The tight binding parameters can be determined by comparing the band structure to muffin-tin results, as we show below.

\section*{Muffin-tin calculations}
The tight binding parameters are unknown for any new system, which presents an obvious difficulty when trying to model experiments. In order to bridge this gap between tight binding description and experimental reality, we make use of the so-called muffin-tin model. To make a good estimate of the tight binding parameters, the muffin-tin approximation is used to calculate the band structure, and the band structure calculated from tight binding is fit to this by tuning the parameters until the band strucutres match. The tight binding LDOS spectra and maps subsequently calculated using these parameters parallel the experimental results quite closely.
 The experimental anti-lattice setup formed by the CO on Cu(111) can be modelled by describing the CO molecules as disk-shaped protrusions in an otherwise constant 2D  potential landscape, resembling an inverted muffin-tin. For the calculations done here, a disk diameter of 0.6 nm (based on observation from STM scans) and a potential height of 0.9 eV were used. 

\noindent In the case of a periodic system, we can use Bloch theory to approximate the band structure of electrons in this periodic potential using the first 5 Fourier components. The resulting band structures can then be used to identify the tight binding parameters (see Supplementary Figure 1, where the muffin-tin band structure and the corresponding tight binding match are shown for both designs used).

\noindent For the higher bands in the tight binding model, the match becomes less accurate due to the interference with p-bands that are not included in the tight binding description, but are present in the muffin-tin model. The obtained parameters are displayed in Supplementary Table 1.
\begin{center}
\begin{table}[h!]
\centering
 \begin{tabular}{||c| c| c| c| c| c| c| c||} 
 \hline
 Design & $t_0$ & $t_1$ & $t_{n_0}$ & $t_{n_1}$ & $s_0$ & $s_1$ & $e$ \\ 
 \hline\hline
 $t_1 < t_0$ & -0.13 eV & 0.7 $t_0$ & 0.2 $t_0$ & 0.02 $t_0$ & 0.2 & 0.12 & -0.105 eV \\ 
 \hline
 $t_1 < t_0$ & 0.67 $t_1$ & -0.13 eV & 0 & 0 & 0.1 & 0.15 & -0.005 eV \\
 \hline

\end{tabular}
 \caption{\doublespacing Tight binding parameters obtained by matching to muffin-tin band structures. Here, $e$ is the on-site energy, $t_0$  and $t_1$ are NN hopping parameters, $t_{n_0}$  and $t_{n_1}$ are NNN hopping parameters, and $s_0$  and $s_1$ are orbital overlaps, each of them intra- and inter-hexagon, respectively.}
 \label{table:1}
\end{table}
\end{center}

\noindent The experimental results (LDOS spectra and maps) were also simulated using the muffin tin model. For a finite system, we numerically solved the non-interacting Schrödinger equation for the muffin-tin potential using von Neumann boundary conditions. After including broadening in the same way as for the tight binding model described above, we find that the muffin-tin results closely match the experimental findings, as shown in Supplementary Figure 2 for the local density of states maps and in Supplementary Figure 3 for the local density of states spectra. In these figures, tight binding results have also been included for comparison.

\section*{Choice of blocking}
At the edge of artificial lattices built by confining the surface state of a metal, there can be significant broadening as a result of the states within the lattice interacting with the surrounding free surface state. In order to observe the edge states corresponding to the finite-size system, it is important to minimize the coupling to the surrounding 2D electron gas. This is achieved by placing additional CO molecules at the boundaries of the structure. However, care should be taken because the positioning of CO molecules outside their regular anti-lattice modulates the on-site energy of neighbouring sites, which leads to a spurious modulation of the spectrum. In order to find suitable positions to place the “blocker” CO molecules, several potential designs were calculated for each of the four lattices using muffin-tin, and those that yielded the best fit to the tight binding predictions were chosen. An example using the molecule zigzag edge for $t_0<t_1$ is shown in Supplementary Figure 4. Here, multiple different blocker positions (and an edge with no blocking) are shown alongside the corresponding muffin-tin spectra. The broadening is plain to see in the LDOS calculated for the design with no blocking (Supplementary Figure 4a) – no clear similarities to the TB-calculated LDOS are seen. When blockers are introduced too far away as in Supplementary Figure 4b, familiar behaviour is observed, except the on-site energy of the edge sites are shifted to lower energies because the edge states are less confined. Upon repositioning the blocker-CO molecules one Cu(111) atom distance closer (Supplementary Figure 4c), the on-site energies of the sites are shifted towards higher energies. Finally in Supplementary Figure 4d, after shifting the blockers one site closer still, the on-site energy becomes approximately comparable to the bulk minimum. Thus, this was the design chosen for our investigation. Similar calculations were performed for the three remaining lattices to find the best position for CO blockers.

\section*{Defects at the edges}
One of the most exciting aspects of topological insulators is the resilience of the edge modes to non-symmetry breaking defects. Here, we investigate the introduction of CO molecules at the edge to behave as defects. Supplementary Figure 5 shows experimental differential conductance maps, where defects have been introduced. Sites on opposite sides of the defect should only couple very weakly via the defect, possibly affecting the shape of the edge. By examining Supplementary Figure 5, it can be seen that the edge modes still exist despite the defects, even in close proximity to them, thus substantiating the topological character of the edge.

\newpage{}
\section*{Supplementary Figures}

\begin{figure}[h!]
    \centering
    \includegraphics[scale=.85]{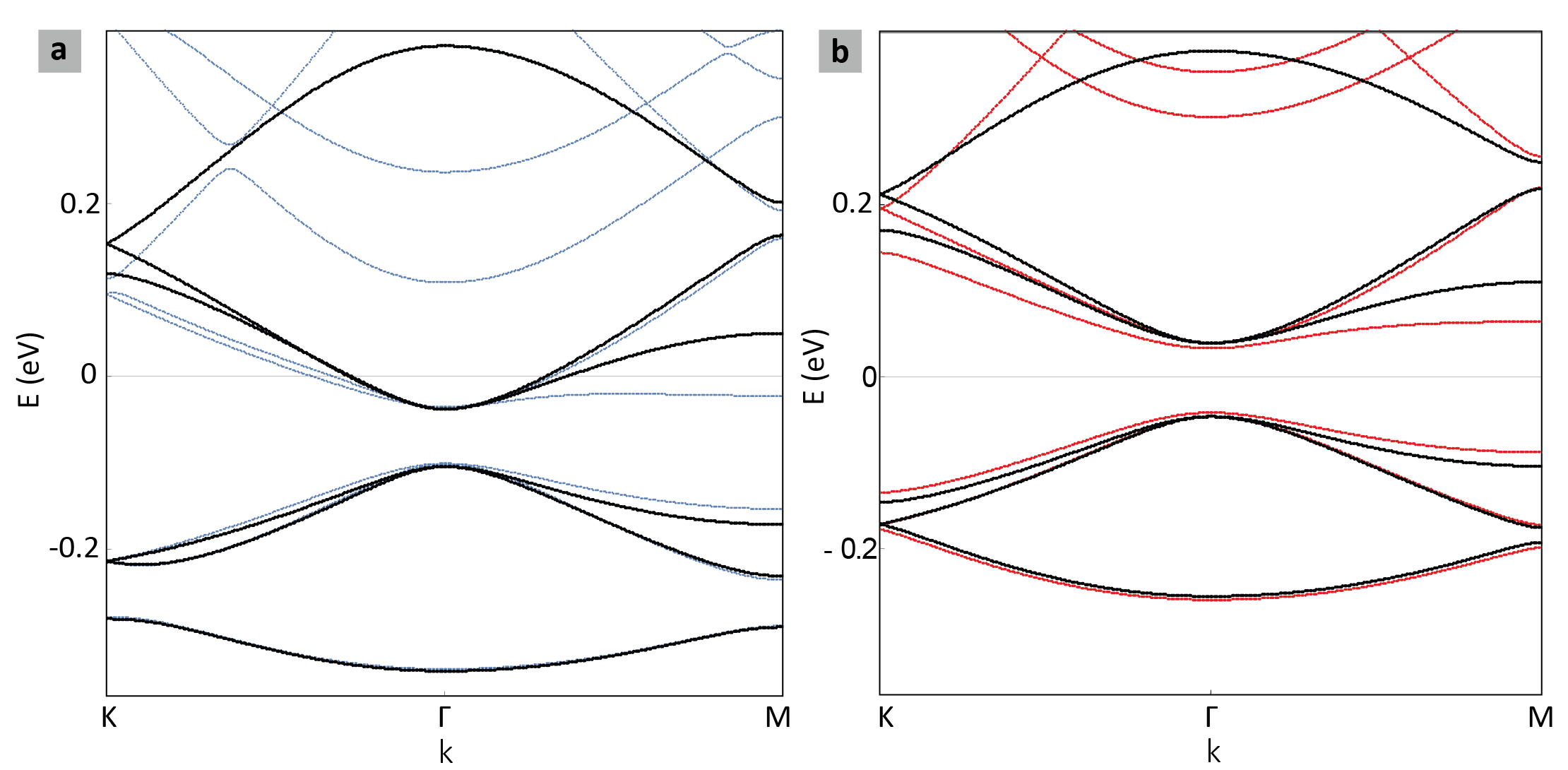}
    \caption{\doublespacing \textbf{ Band structure for the periodic Kekulé lattice}. \textbf{a} $t_1 < t_0$ and \textbf{b} $t_0 < t_1$. The tight binding fit is displayed in black, and the muffin-tin band structure in colour (blue and red, respectively).}
\end{figure}

\begin{figure}[h!]
    \centering
    \includegraphics[scale=.85]{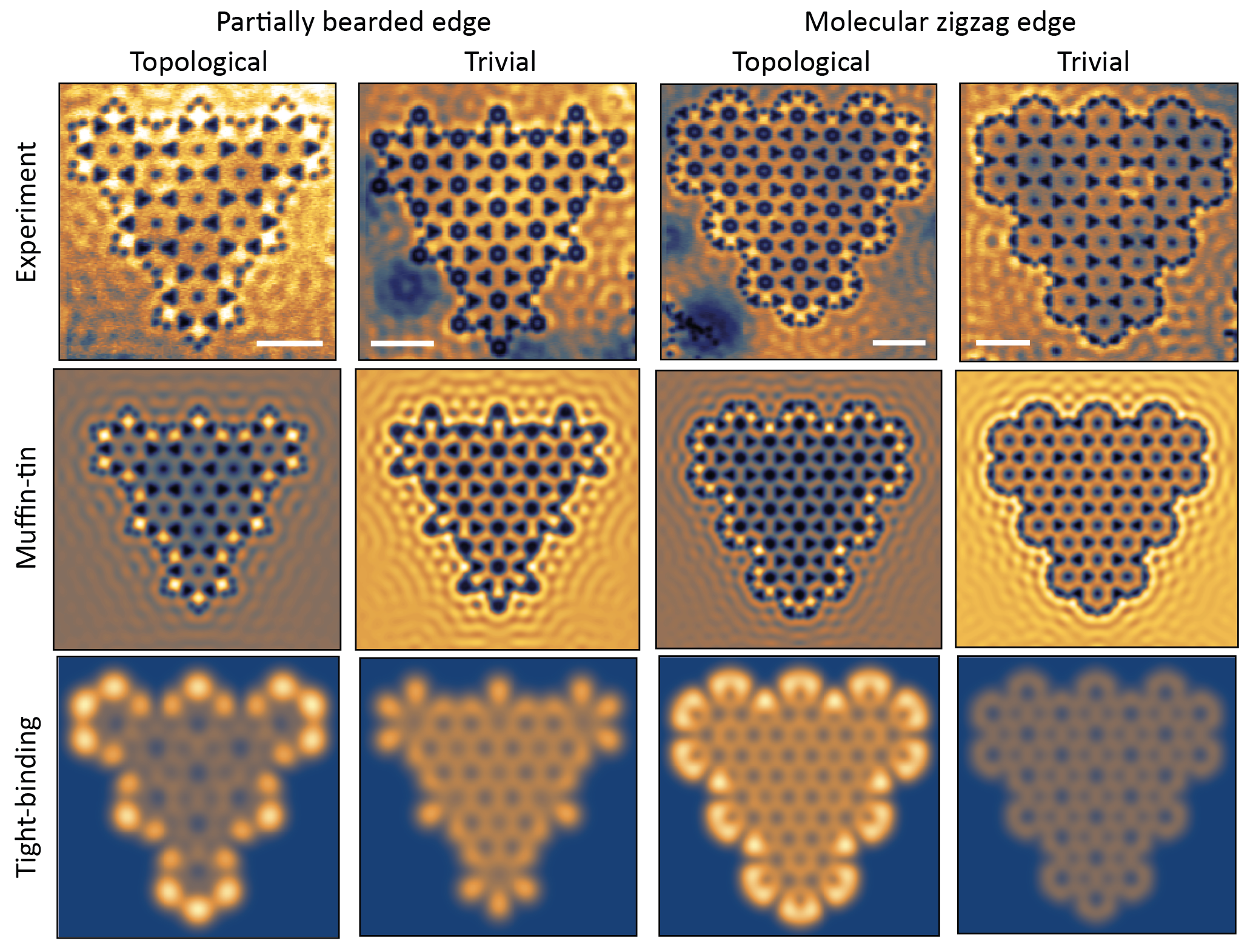}
    \caption{\doublespacing \textbf{Comparison of LDOS maps.} Local density of states maps at the gap energy for both edge types obtained experimentally (top row), with muffin-tin (centre row) and with tight binding (bottom row). Scale bars in white indicate 5 nm.}
\end{figure}

\begin{figure}[h!]
    \centering
    \includegraphics[scale=1]{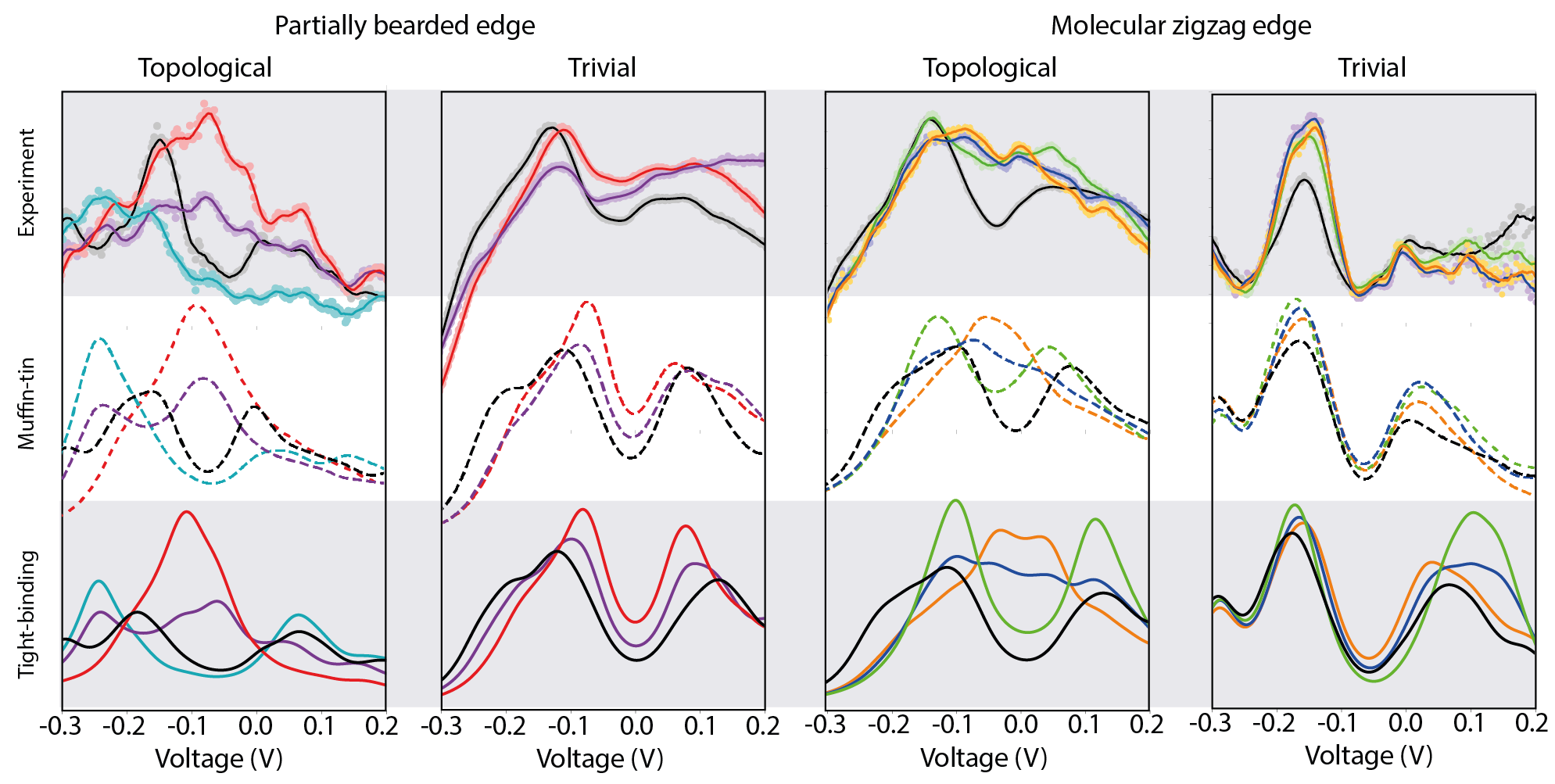}
    \caption{\doublespacing \textbf{Comparison of LDOS spectra.} Local density of states spectra for both edge types obtained experimentally (top row), theoretically using a muffin-tin calculation (middle row) and with tight binding (bottom row). Each colour refers to a type of site, as illustrated in Fig. 1 of the main text.}
\end{figure}

\begin{figure}[h!]
    \centering
    \includegraphics[scale=1.5]{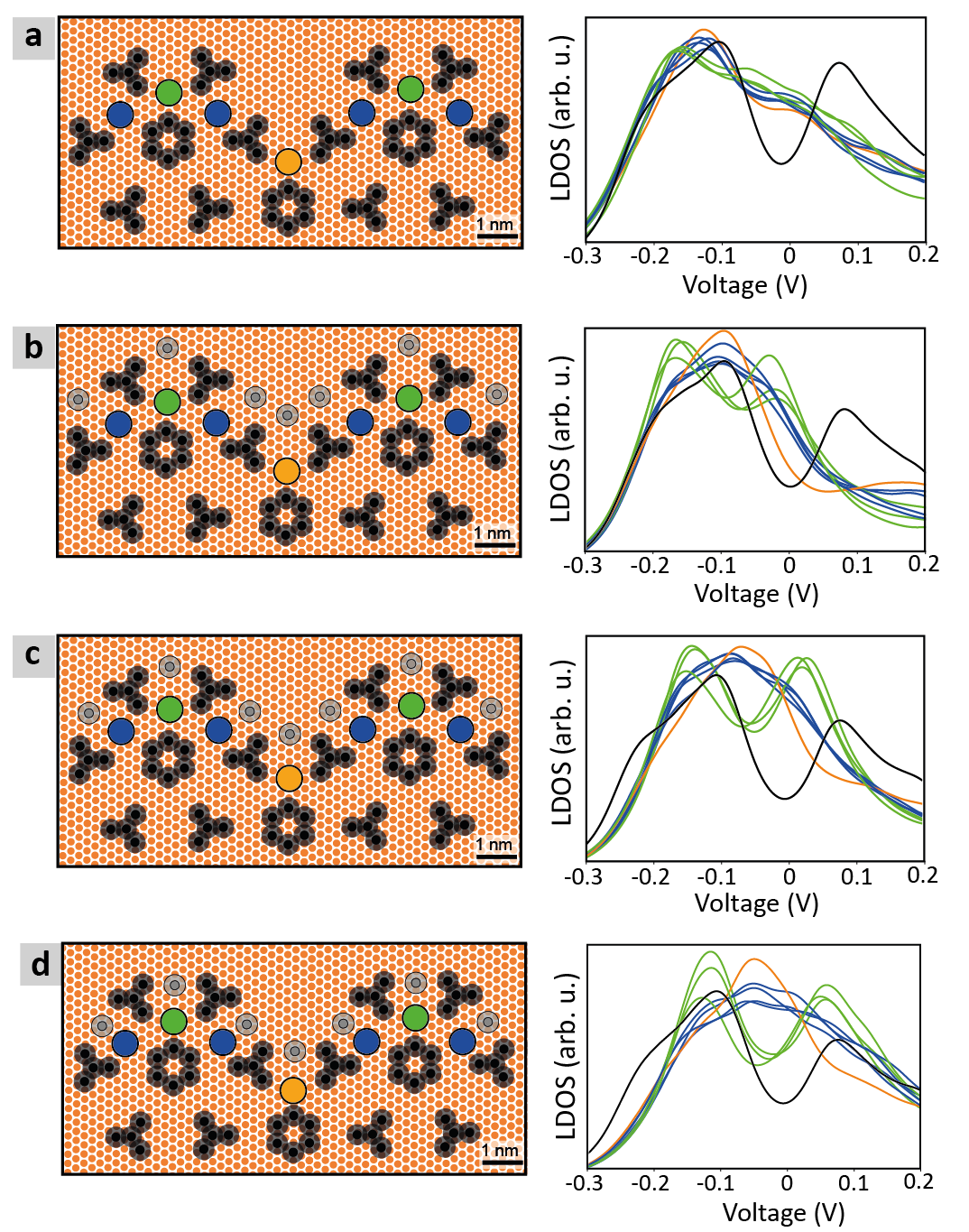}
    \caption{\doublespacing \textbf{The effect of protective “blocker” CO molecules.} The left column shows positions of CO molecules (black circles) on the Cu(111) surface (orange dots) to form the molecule zigzag edge of the Kekulé lattice. Grey dots represent CO molecules that are used to shield the electronic states in the lattice from unwanted interactions with the surrounding 2DEG. The spectra were calculated on the dots of corresponding colour, except the black line, which was calculated for the bulk. A spectrum was taken for equivalent sites at the corner and edges, thus there are multiple spectra of each colour. \textbf{a} The lattice with no blocking CO molecules and the corresponding spectra. \textbf{b} Blocking at distant positions, leading to a shift of the edge modes to lower energy. \textbf{c} An improvement on \textbf{b}, where the blocking CO molecules are shifted one Cu(111) atom distance closer. \textbf{d} The final design used in our investigation, where the CO molecules are pushed one additional site closer and the on-site energy of the edge mode is comparable to the energy of the bulk gap.}
\end{figure}

\begin{figure}[h!]
    \centering
    \includegraphics[scale=1]{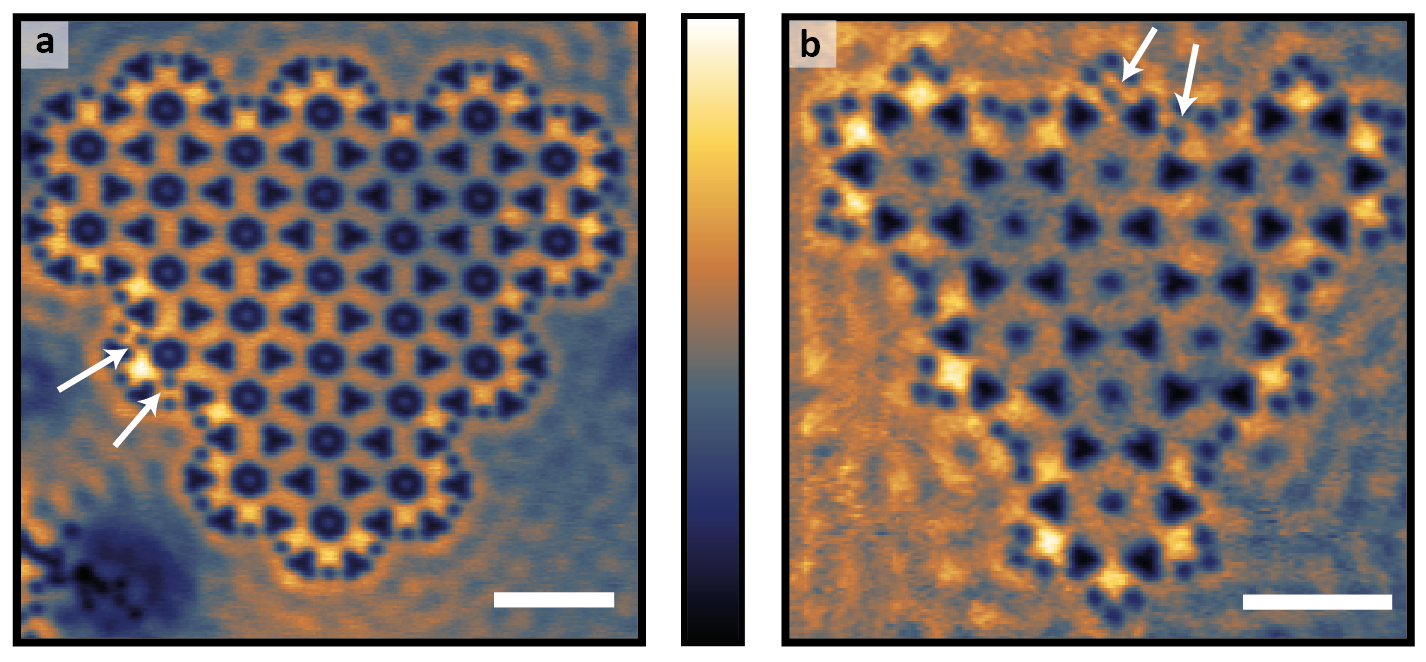}
    \caption{\doublespacing \textbf{Defects in edges.} Experimental differential conductance maps showing the effect of defects in the edge modes of the lattices. Brighter pixels represent higher LDOS. White arrows point to the defects. Scale bars represent 5 nm.  \textbf{a} The molecular zigzag edge with $t_0<t_1$  (acquired at -40mV). \textbf{b} The partially bearded edge with $t_1<t_0$ (acquired at -65mV).}
\end{figure}